\documentclass[%
 reprint,
 superscriptaddress,
 amsmath,amssymb,
 aps,
 prl,
]{revtex4-2}

\usepackage{graphicx}
\usepackage{dcolumn}
\usepackage{bm}
\usepackage{color}
 \usepackage{ulem}

\begin{document}

\preprint{APS/123-QED}

\title{Continuous-Variable Nonclassicality Detection under Coarse-Grained Measurement}

\author{Chan Roh}
\affiliation{Department of Physics, Korea Advanced Institute of Science and Technology, Daejeon 34141, Korea}
\author{Young-Do Yoon}
\affiliation{Department of Physics, Korea Advanced Institute of Science and Technology, Daejeon 34141, Korea}
\author{Jiyong Park}
 \email{jiyong.park@hanbat.ac.kr}
\affiliation{School of Basic Sciences, Hanbat National University, Daejeon 34158, Korea}
\author{Young-Sik Ra}
 \email{youngsikra@gmail.com}
\affiliation{Department of Physics, Korea Advanced Institute of Science and Technology, Daejeon 34141, Korea}

\date{\today}

\begin{abstract}
Coarse graining is a common imperfection of realistic quantum measurement, obstructing the direct observation of quantum features. Under highly coarse-grained measurement, we experimentally detect the continuous-variable nonclassicality of both Gaussian and non-Gaussian states. Remarkably, we find that this coarse-grained measurement outperforms the conventional fine-grained measurement for nonclassicality detection: it detects nonclassicality beyond the reach of the variance criterion, and furthermore, it exhibits stronger statistical significance than the high-order moments method. Our work shows the usefulness of coarse-grained measurement by providing a reliable and efficient way of nonclassicality detection for quantum technologies.
\end{abstract}

\maketitle



A continuous-variable (CV) quantum state of light is a versatile quantum resource for quantum information technologies: quantum states encoded in the continuous field quadratures are used for quantum computing~\cite{Larsen19,Asavanant19,Ra20,Zhong:2021ec,Madsen:2022jm,Pfister:2019ck}, quantum communication~\cite{Zhang:2020jv,Kovalenko:2021ej}, and quantum metrology~\cite{Guo20,Nielsen:2021tt}. In particular, even a single-mode CV quantum state---where entanglement is absent by construction---can show quantum enhancement, e.g., in quantum parameter estimation~\cite{Nielsen:2021tt,Oh:2019ek} and quantum key distribution~\cite{Derkach:2020kl}. Furthermore, a single-mode CV quantum state (e.g., a squeezed state and a Gottesman-Kitaev-Preskill state~\cite{GKP}) is a basic building block to construct a large-scale entangled system~\cite{Larsen19,Asavanant19,Kouadou:2022hj,Kim02,Asboth05}. To exploit such quantum resources in practice, it is essential to verify the nonclassicality of experimentally generated states.

A nonclassical state is defined as a state which cannot be expressed as a statistical mixture of coherent states~\cite{Asboth05}. Nonclassicality can be in principle detected by performing quantum state tomography, but it is a demanding process requiring informationally-complete measurements and maximum-likelihood reconstruction~\cite{DAriano:1994iy,Lvovsky:2009fr}. Alternatively, measuring the variance of a single quadrature alone can detect nonclassicality~\cite{Andersen:2016fu}; however, its application is generally limited to Gaussian states. For detecting the nonclassicality of non-Gaussian states, there are methods of measuring high-order moments~\cite{Agarwal93,Shchukin05} or a characteristic function~\cite{Vogel00,Kiesel09}, but these require substantial data collection to achieve sufficient statistical significance. Even worse, under \textit{coarse-grained measurement}, all of these methods are subject to \textit{false-positive} detection of nonclassicality, and thus, careful considerations must be made~\cite{Schneeloch13,Tasca13,Park14}.

Coarse graining commonly occurs in realistic quantum measurement~\cite{Kofler:2007bd}, where nearby measurement outcomes are grouped together as a single bin, thereby producing the same result. It originates from a finite resolution in measurement, for example, when using image pixels~\cite{Tasca13} and quadrature~\cite{Gabriel:2010gb} and photon-number~\cite{Raeisi:2011kw} detection. Coarse graining, like decoherence processes, makes it hard to observe quantum features by inducing a quantum-to-classical transition~\cite{Raeisi:2011kw}. In dealing with realistic situations, it is, therefore, necessary to establish reliable nonclassicality criteria compatible with coarse-grained measurement.

In this Letter, we demonstrate reliable detection of CV nonclassicality, even under considerable coarse graining in measurement. In the experiment, we show that the nonclassicality of squeezed vacuum states is directly detected under coarse-grained quadrature measurement, which is obtained only at a single quadrature (namely, the $\hat{x}$ quadrature). Interestingly, our method based on coarse-grained measurement can detect the nonclassicality of a non-Gaussian state of a phase-diffused squeezed vacuum, where the conventional variance measurement---even without coarse graining---fails to detect~\cite{Andersen:2016fu}. Furthermore, the simplicity of our method, requiring only multiplication and division of experimental data, considerably reduces the sampling error for nonclassicality detection. As a result, our method shows better performance than the conventional moments method (requiring matrix decompositions~\cite{Agarwal93,Shchukin05}) by providing stronger statistical significance on nonclassicality detection.

Let us start by explaining our nonclassicality test for CV quantum states. We consider a single-quadrature probability distribution $p(x)$ of a quantum state, i.e., a marginal distribution of a Wigner function. We construct a nonclassicality test by noticing that the probability distribution of any classical state cannot exhibit a narrower structure than that of a coherent state; it is because a classical state should be expressed by a statistical sum of coherent states. Consequently, a nonclassicality test can be formulated by comparing widths of probability distributions by a given state and a coherent state:
\begin{equation}
    R(s) = {p(s)p(-s) \over p(0)^2} e^{s^2},
    \label{eq:three_point_test}
\end{equation}
where $R<1$ certifies nonclassicality~\cite{Park21}. We choose three points of $x$ $\in (-s, 0, s)$, where $s>0$, for the test, which is favorable for a probability distribution having a peak at the origin, but the method can be generalized to detect nonclassicality of generic quantum states~\cite{Park21}.
In our convention, the bosonic commutation relation of quadrature operators $\hat{x}$ and $\hat{p}$ is given by $[ \hat{x}, \hat{p} ] = 2 i$. The nonclassicality test in Eq.~(\ref{eq:three_point_test}) can be adapted for coarse-grained measurement~\cite{Park21}:
\begin{equation}
    \mathcal{R} = {C_{d}C_{-d} \over (C_0)^2}~e^{\sigma^2 d^2}.
    \label{eq:three_bin_test}
\end{equation}
$C_m$ represents the count of measurement outcomes in a bin index of an integer $m$, which has a range of $x = [(m-1/2)\sigma,(m+1/2)\sigma]$ with a bin size $\sigma$. Similar to Eq.~(\ref{eq:three_point_test}), three indices of $m$ $\in (-d, 0, d)$ are chosen, where $d$ is a positive integer. $\mathcal{R}<1$ certifies the nonclassicality of a given state, which is applicable for both Gaussian states and non-Gaussian states~\cite{Park21}. Note that $\mathcal{R}$ can be estimated from simple multiplication and division of obtained counts $C_m$ together with a predetermined value $e^{\sigma^2 d^2}$; the estimation error of $\mathcal{R}$ is thus attributed to the statistics of $C_m$ only. We call this non-classicality test, Eq.~(\ref{eq:three_bin_test}), a \textit{three-bin test}.

\begin{figure}[t]
    \centering
    \includegraphics[width = 85mm]{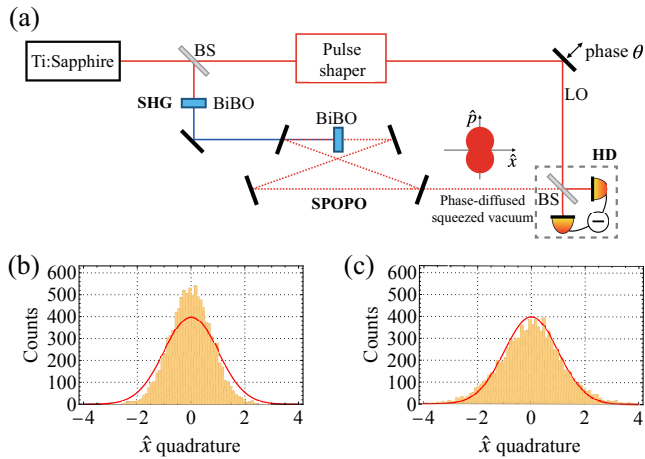}
    \caption{(a) Experimental setup. Ti:Sapphire laser produces femtosecond pulses, which are used for second harmonic generation (SHG) and local oscillator (LO). Homodyne detection (HD) measures squeezed light generated from a synchronously pumped optical parametric oscillator (SPOPO).
    (b) Histogram of $\hat{x}$ quadrature outcomes by a squeezed state; the variance is $-2.3 \pm 0.1$ dB. (c) Histogram of $\hat{x}$ quadrature outcomes by a phase-diffused squeezed state; the variance is $0.23 \pm 0.07$ dB, being larger than the vacuum variance.
The data number is $10^4$ for each of (b,c). The red line is the vacuum distribution as a reference.}
    \label{fig:phase_diffusing}
\end{figure}

We experimentally demonstrate the three-bin test on squeezed vacuum states with various phase diffusions. Figure~\ref{fig:phase_diffusing}(a) describes the experimental setup. A mode-locked Ti:Sapphire laser produces a 75-fs pulse train with a repetition rate of 80 MHz and a central wavelength of 800 nm. In the second harmonic generation, a pulse laser of 400 nm wavelength is generated, which is used for the pump of SPOPO. SPOPO has a free spectral range of 80 MHz to match the repetition rate of the Ti:Sapphire laser, its finesse is 27.4, and it contains a 2-mm-thick BiBO crystal for type-I spontaneous parametric down-conversion. SPOPO generates squeezed light as operating below its threshold~\cite{Ra20,Cai17}.

To measure the generated squeezed light, we employ homodyne detection shown in Fig.~\ref{fig:phase_diffusing}(a). A local oscillator (LO) beam, determining the mode of homodyne detection, is engineered by a pulse shaper for spectral mode matching with SPOPO~\cite{Ra20,Cai17}. The visibility and the clearance of the homodyne detection are 95 \% and 15 dB, respectively. We measure two sideband frequencies simultaneously (1 MHz and 2 MHz with a sampling rate of 100 kHz for each): the former is for obtaining a quadrature outcome $x$, and the latter is for phase information $\theta$. As varying the phase $\theta$ by a piezoelectric transducer, we obtain a pair of data $(\theta_i, x_i)$ for each measurement $i$. For choosing $\hat{x}$ quadrature measurement, we select data within a small range of phase $\theta_i \in (-0.087,0.087)$ rad.


Figure ~\ref{fig:phase_diffusing}(b) shows the obtained outcomes for $\hat{x}$ quadrature. Compared with the vacuum state, the squeezed state shows a narrower distribution, resulting in $-2.3 \pm 0.1$ dB variance. We further characterize the amount of phase diffusion present in the generated state. For this purpose, we use the variance and the kurtosis of $\hat{x}$ quadrature measurement and the variance of $\hat{p}$ quadrature measurement (see Sec. I of the Supplemental Material for details~\cite{SM}). The estimated phase diffusion is $\Delta_0 = 0.15 \pm 0.02$ rad, which originates from interferometer instability and the phase estimation noise. To increase the phase diffusion further, we add a random noise in a normal distribution $\epsilon \sim \mathcal{N} (0,\Delta_e^2)$ to the estimated phase $\theta_i$:
\begin{equation}
    (\theta_i,x_i) \rightarrow (\theta_i+\epsilon, x_i),
\end{equation}
which increases the phase diffusion to $\Delta =  \sqrt{\Delta_0^2+\Delta_e^2}$. After adding the phase noise, we select data in the same way as before for $\hat{x}$ quadrature measurement. We have confirmed that the resulting phase diffusion agrees well with the prediction (see~\cite{SM}). Figure~\ref{fig:phase_diffusing}(c) shows the obtained quadrature outcomes, resulting in a variance of $0.23 \pm 0.07$ dB (i.e., no squeezing) and a phase diffusion of $0.37 \pm 0.01$ rad. Note that this phase-diffused state is still nonclassical while the variance criterion ($\langle \delta \hat{x}^2 \rangle < 1$) fails to detect its nonclassicality.

\begin{figure*}[t]
    \centering
    \includegraphics[width = 170mm]{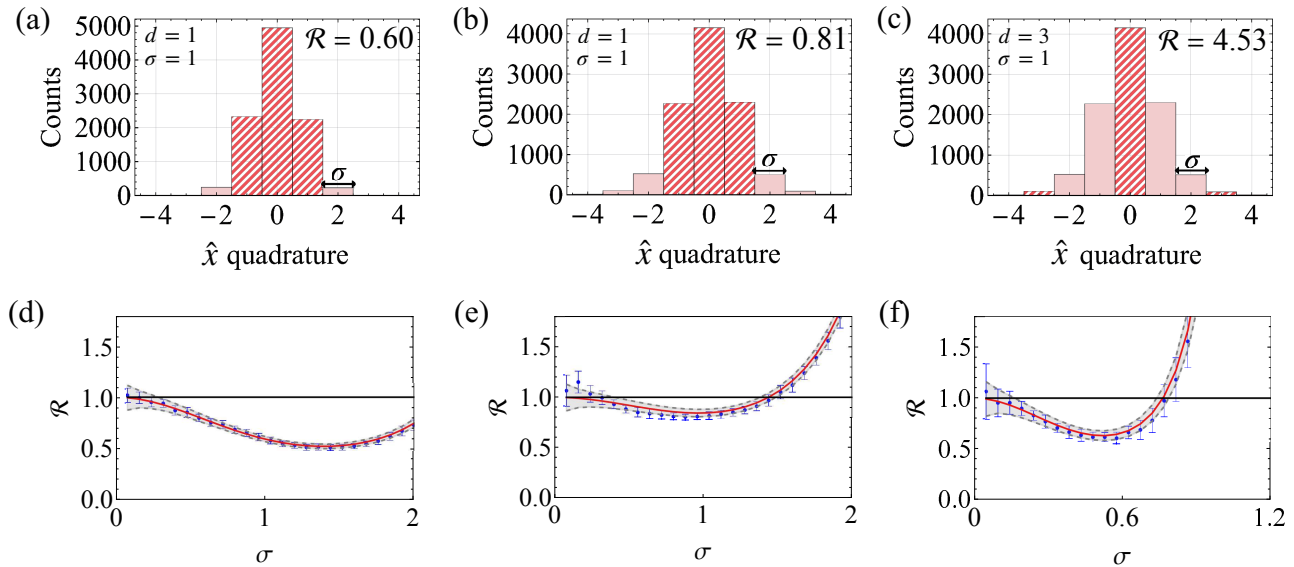}
    \caption{Nonclassicality detection with coarse-grained quadrature measurement. Quadrature outcomes (total number: $10^4$) are coarse-grained with a bin size $\sigma$ of one: (a) the squeezed state in Fig.~\ref{fig:phase_diffusing}(b); (b,c) the phase-diffused state in Fig.~\ref{fig:phase_diffusing}(c). For the nonclassicality detection using Eq. (\ref{eq:three_bin_test}), the three hatched bins are selected ($d=1$ for (a,b), and $d=3$ for (c)). (d-f) Effect of the bin size on nonclassicality detection for (a-c), respectively. Blue dots are experimental data with error bars of one standard deviations. To obtain the statistics, we use the bootstrap method by sampling $10^4$ data from the total $4 \times 10^4$ experimental data. Red lines are theoretical curves, where the shaded areas represent statistical errors by considering a finite data number ($10^4$). Black horizontal lines are the thresholds for detecting nonclassicality (nonclassical if $\mathcal{R}<1$).}
    \label{fig:three_bin}
\end{figure*}

Now we consider coarse-grained quadrature measurement. As the first example, we make binning on the data for the squeezed state in Fig.~\ref{fig:phase_diffusing}(b). The result is shown in Fig.~\ref{fig:three_bin}(a) where the bin size $\sigma$ is 1. The three-bin test with $d=1$ detects the nonclassicality of the state under coarse-grained measurement, showing $\mathcal{R} = 0.60 \pm 0.04 < 1$. Next, we investigate the phase-diffused squeezed state in Fig.~\ref{fig:phase_diffusing}(c), where the variance criterion fails on nonclassicality detection. Figure~\ref{fig:three_bin}(b) shows the result of coarse graining with $\sigma=1$. By conducting the three-bin test, we again find a clear evidence of nonclassicality, $\mathcal{R} = 0.81\pm0.04 < 1$. The three-bin test successfully detects the nonclassicality of both the squeezed state and the phase-diffused state.

We further investigate the effects of the bin size and the bin distance on the nonclassicality detection. For the squeezed state, we find that the three-bin test works for a wide range of bin size (Fig.~\ref{fig:three_bin}(d)), where $\sigma=1.4$ is close to the optimum, showing $\mathcal{R} = 0.51 \pm 0.03 < 1$. The increase of $\mathcal{R}$ for a large $\sigma$ is due to a substantial coarse-graining effect, and for a small $\sigma$, the standard deviation of $\mathcal{R}$ increases due to the limited number of counts collected in a single bin. For the phase-diffused squeezed state, we also find a similar behavior, as shown in Fig.~\ref{fig:three_bin}(e). To investigate the bin distance effect, we perform the three-bin test by increasing the distance to $d=3$ for the phase-diffused squeezed state, as shown in Fig.~\ref{fig:three_bin}(c). In this case, $\mathcal{R} = 4.53 \pm 0.74 > 1$ for $\sigma = 1$, but we again detect nonclassicality by decreasing $\sigma$ (Fig.~\ref{fig:three_bin}(f)), where even a larger deviation from the classical limit ($\mathcal{R}=1$) is found compared with Fig.~\ref{fig:three_bin}(e): $\mathcal{R} = 0.62 \pm 0.05< 1$ at $\sigma = 0.5$. The three-bin test can detect nonclassicality with a wide range of $\sigma$ without elaborate optimization.


Let us compare the performance of the three-bin test with the conventional moment method (even without coarse graining)~\cite{Agarwal93,Shchukin05}. In the moment method, an $n \times n$ correlation matrix is constructed by up to $n$-th order moments (for detailed explanations, refer to Sec. II of~\cite{SM}). If its smallest eigenvalue is negative ($\lambda(n) < 0$), then the given state is nonclassical; in principle, this method can detect nonclassicality for any quantum state if a sufficiently large $n$ with an infinite number of data is available~\cite{Agarwal93}. In practice, however, because the number of data is finite, the standard deviation ($\delta\lambda(n)$) increases rapidly as $n$ increases, which makes it difficult to attain sufficient statistical significance. To quantitatively compare the statistical significances of the three-bin test (bin) and the moment method (moment), we introduce a violation degree $\mathcal{V}$, which is defined as the ratio between the distance from the classical limit and its standard deviation:
\begin{eqnarray}
    \mathcal{V}_\mathrm{bin} &=& {1-\bar{\mathcal{R}} \over \delta \mathcal{R}} \nonumber \\
    \mathcal{V}_\mathrm{moment}(n) &=& {-\bar{\lambda} (n)\over \delta \lambda (n)}.
    \label{eq:violation}
\end{eqnarray}
The upper bar and $\delta$ denote the mean value and the standard deviation, respectively. The classical limits of 1 and 0 have been used for $\mathcal{R}$ and $\lambda (n)$, respectively.
A positive (negative) $\mathcal{V}$ shows detection (no detection) of nonclassicality, and the larger $\mathcal{V}$ indicates the stronger statistical significance of nonclassicality detection.

\begin{figure}[t]
    \centering
    \includegraphics[width = 100mm]{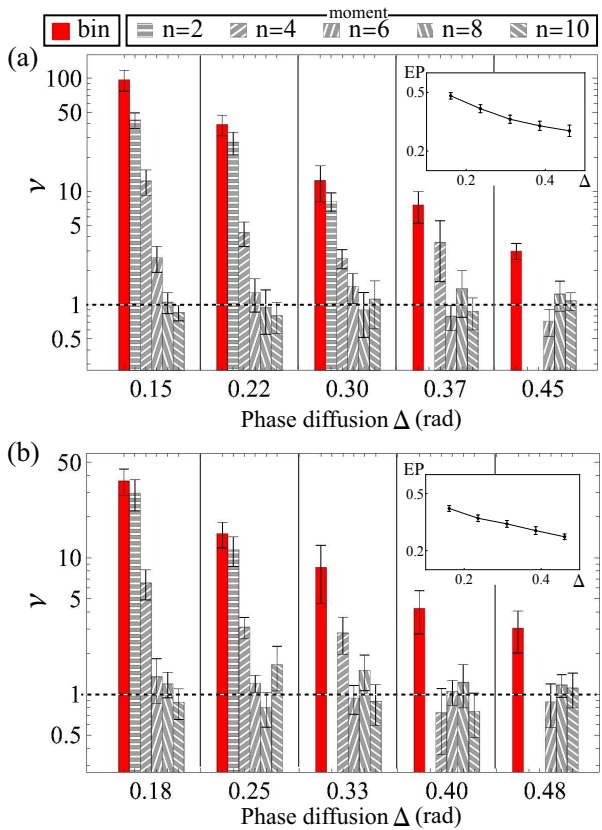}
    \caption{Performance comparison between the three bin test (bin) and the moment method (moment).
In (a), the initial state has ($\hat{x}$, $\hat{p}$) quadrature variances of ($-2.3 \pm 0.1$, $7.0 \pm 0.1$) dB with $\Delta = 0.15 \pm 0.02$ rad, and in (b), the corresponding conditions are ($-1.6 \pm 0.1 $, $7.0 \pm 0.1$) dB with $\Delta = 0.18 \pm 0.02$ rad. The violation degree $\mathcal{V}$ quantifies the statistical significance of nonclassicality detection, as defined in Eq.~(\ref{eq:violation}).
The horizontal dashed lines represent $\mathcal{V}=1$ where the standard deviation is as large as the mean value to detect nonclassicality. When $\mathcal{V}$ is negative (i.e., a failure of nonclassicality detection), the corresponding plot is omitted for clarity. The insets plot the entanglement potential (EP) of the states under investigation. Error bars, representing one standard deviation, are obtained from five repeated experiments.
}
    \label{fig:performance}
\end{figure}

Figure~\ref{fig:performance} compares the violation degrees of the two methods, $\mathcal{V}_{\mathrm{bin}}$ and $\mathcal{V}_{\mathrm{moment}}$, tested for initial squeezing of (a) $-2.3 \pm 0.1$ dB and (b) $-1.6 \pm 0.1$ dB. The moment method by $n=2$ is equivalent to the variance criterion: $\mathcal{V}_{\mathrm{moment}}(2)>0$ leads to $\langle\delta \hat{x}^2 \rangle < 1$. One can find in Fig.~\ref{fig:performance} that, for a large phase diffusion $\Delta \ge 0.37$, the moment method at $n=2$ cannot detect nonclassicality ($\mathcal{V}_{\mathrm{moment}}(2) < 0$, thus omitted in the figure), which agrees with the previous discussion for Fig.~\ref{fig:phase_diffusing}(c). Higher order moments ($n \ge 4$) can detect the nonclassicality of such non-Gaussian states; however, for a larger $n$, due to the increased standard deviation $\delta \lambda (n)$, $\mathcal{V}_{\mathrm{moment}}$ generally decreases as shown in Fig.~\ref{fig:performance}. On the other hand, the three-bin test reliably detects nonclassicality in the entire range of phase diffusions shown, outperforming the moment method. The three-bin test becomes more powerful than the moment method for a larger phase diffusion and squeezing.

Here we further examine the operational significance of the phase-diffused squeezed state (naturally arising by optical propagation) as quantum resources. We place two insets in Fig.~\ref{fig:performance} to display the dynamics of the squeezed vacuum's entanglement potential~\cite{Asboth05} under phase diffusion. The entanglement potential of a quantum state quantifies the amount of quantum entanglement producible by injecting the quantum state through a passive beam splitter. Notably, such quantum entanglement is essential in multipartite quantum information tasks~\cite{Hage:2008fl,Loock00}. The insets exhibit that the entanglement potential remains positive under strong phase diffusions, which reveals the complex and robust nature of CV quantum resources beyond quantum squeezing.


In conclusion, we have experimentally demonstrated a reliable CV nonclassicality test robust under coarse-grained measurement, i.e., the three-bin test. This test employs coarse-grained data from single-quadrature measurement, which directly captures a phase-space structure narrower than the vacuum fluctuation, thereby showing no false detection of nonclassicality under coarse-graining. We have tested phase-diffused squeezed vacuum to compare the performances of our three-bin test and the conventional moment method. In our experiments, the three-bin test outperformed the other in the robustness under phase diffusion and the statistical significance. It is remarkable because the three-bin test does not require fine-grained measurements and complex data processings. We have also addressed the operational relevance of the witnessed nonclassicality by examining the entanglement potential. Such nonclassicality can also be converted to quantum squeezing~\cite{Hage07, Filip13}.

Our results strongly suggest that systematic and rigorous approaches to coarse-graining models may provide fundamental and practical tools in quantum information technologies. We expect that our contributions will facilitate future studies to uncover the rich structure of CV quantum resources. For instance, it may be interesting to test quantum non-Gaussianity~\cite{Filip11, Lachman22} with coarse-grained data, e.g., three bins, by using energy information~\cite{Genoni13, Hughes14} or more quadratures~\cite{Happ18}.

\begin{acknowledgments}
This work was supported by the Ministry of Science and ICT (NRF-2020M3E4A1080028, NRF-2022R1A2C2006179, NRF-2019R1G1A1002337), and MSIT of Korea under the ITRC support program (IITP-2022-2020-0-01606).
\end{acknowledgments}

\end{document}


\preprint{}

\title{Supplementary Materials for `Continuous-Variable Nonclassicality Detection under Coarse-Grained Measurement'}

\author{Chan Roh}
\affiliation{Department of Physics, Korea Advanced Institute of Science and Technology, Daejeon 34141, Korea}
\author{Young-Do Yoon}
\affiliation{Department of Physics, Korea Advanced Institute of Science and Technology, Daejeon 34141, Korea}
\author{Jiyong Park}
\email{jiyong.park@hanbat.ac.kr}
\affiliation{School of Basic Sciences, Hanbat National University, Daejeon 34158, Korea}
\author{Young-Sik Ra}
\email{youngsikra@gmail.com}
\affiliation{Department of Physics, Korea Advanced Institute of Science and Technology, Daejeon 34141, Korea}
 
\date{\today}

\maketitle

\section{Measuring the phase diffusion from homodyne data}
We explain how we jointly determine the initial squeezing parameter $r$, optical loss $l$, and phase diffusion $\Delta$ in the experiment.
The initial quantum state is a pure $x$-squeezed vacuum with squeezing variances of $(\delta \hat{x}_0)^2 = e^{-2r}$ and $(\delta \hat{p}_0)^2 = e^{2r}$, where $r>0$. An optical loss is modeled as a beam-splitting interaction between a quantum state and a vacuum state with a reflectance of $l$. The phase diffusion is described as an incoherent mixing of phase rotation to a quantum state $\hat{\rho}$:
\begin{equation}
    \hat{\rho} \mapsto \int_{-\infty}^{\infty} d\theta \frac{e^{-\theta^{2}/2\Delta^{2}}}{\sqrt{2\pi}\Delta} e^{i \hat{n} \theta} \hat{\rho} e^{-i \hat{n} \theta},
\end{equation}
where $\Delta$ determines the amount of phase diffusion. Let us investigate the variance of squeezing and anti-squeezing quadrature, i.e., $\langle (\delta \hat{x})^2 \rangle$ and $\langle (\delta \hat{p})^2 \rangle$, for an $x$-squeezed vacuum under the presence of optical loss and phase noise. We obtain
\begin{equation}
    \begin{aligned}
        \langle (\delta \hat{x})^2 \rangle &= \int_{-\infty}^{\infty} d \theta \;\mathrm{Var}(r,\theta,l) {e^{-\theta^2/2\Delta^2} \over  \sqrt{2 \pi} \Delta} \\
        &= l+(1-l)e^{-\Delta^2}(e^{-2 r} \cosh \Delta^2 +e^{2r} \sinh \Delta^2),
    \end{aligned}
    \label{eq:xvar}
\end{equation}
and
\begin{equation}
    \begin{aligned}
        \langle (\delta \hat{p})^2 \rangle &= \int_{-\infty}^{\infty} d \theta \;\mathrm{Var}(r,\theta+{\pi \over 2},l) {e^{-\theta^2/2\Delta^2} \over  \sqrt{2 \pi} \Delta} \\
        &= l+(1-l)e^{-\Delta^2}(e^{2 r} \cosh \Delta^2 +e^{-2r} \sinh \Delta^2),
    \end{aligned}
    \label{eq:pvar}
\end{equation}
where $\delta \hat{o} = \hat{o}-\langle \hat{o} \rangle$ denotes the deviation operator for $\hat{o}$, and $\mathrm{Var}(r,\theta,l) = l+(1-l)(e^{-2r}\cos^2\theta +e^{2r}\sin^2\theta)$ is the variance of a rotated quadrature $\hat{x} \cos \theta + \hat{p} \sin \theta$ for the $x$-squeezed vacuum under the presence of optical loss only. Similarly, we derive the kurtosis $K$ of the squeezing quadrature as
\begin{equation}
    \begin{aligned}
        K & \equiv {\langle (\delta \hat{x})^4 \rangle \over \langle (\delta \hat{x})^2 \rangle^2} = {{3 \over \langle (\delta \hat{x})^2 \rangle^2} \int_{-\infty}^{\infty} d\theta\; [ \mathrm{Var}(r,\theta,l) ]^2{e^{-\theta^2/2\Delta^2} \over  \sqrt{2 \pi} \Delta}} \\
        &= 3 + { 6 (1-l)^2e^{-4\Delta^2} \sinh^{2} (2\Delta^2) \sinh^{2} (2r) \over \langle (\delta \hat{x})^2 \rangle^2},
    \end{aligned}
    \label{eq:Kurtosis}
\end{equation}
where we have used that kurtosis of a normal distribution is always three \cite{Cramer1946}. In the experiment, we obtain the variances $\langle (\delta \hat{x})^2 \rangle$ and $\langle (\delta \hat{p})^2 \rangle$, and the kurtosis $K$ using the measured homodyne outcomes. We can then numerically calculate the squeezing parameter $r$, optical loss $l$, and phase diffusion amplitude $\Delta$ using Eqs. \eqref{eq:xvar}-\eqref{eq:Kurtosis}.

\begin{figure}
    \centering
    \includegraphics[width = 160mm]{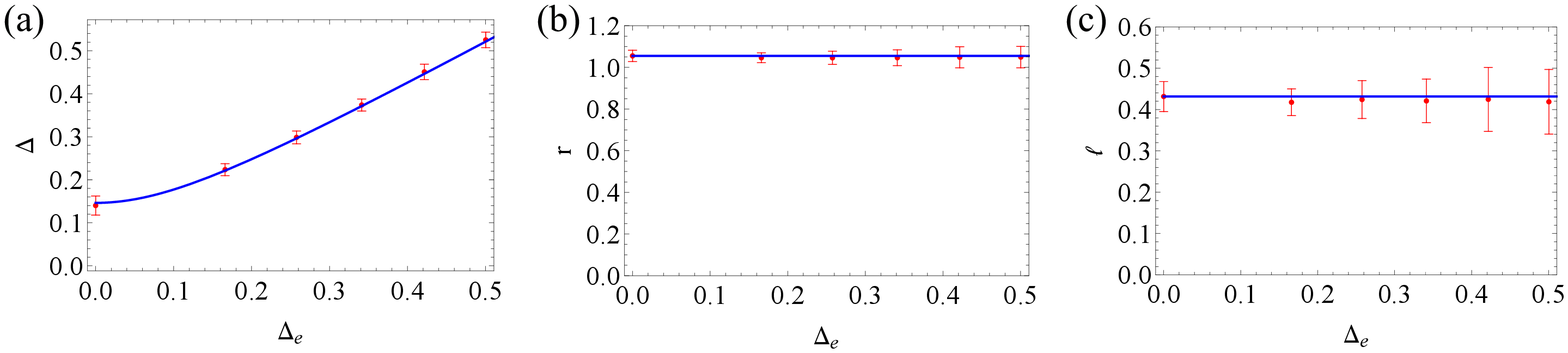}
    \caption{Estimated parameters of (a) the phase diffusion $\Delta$, (b) initial squeezing $r$, and (c) total optical loss $l$. $\Delta_e$ is the standard deviation of an added Gaussian phase noise. Red points are estimated parameters from quadrature data of phase-diffused squeezed vacuum, and blue lines are expected theoretical values. Error bars are obtained by the bootstrap method.}
    \label{fig:diffusing_parameters}
\end{figure}

In the experiment, we collected 40,000 $\hat{x}$ (squeezing) quadrature and $\hat{p}$ (anti-squeezing) quadrature data of each phase-diffused squeezed vacuum and generated 10 sets of 10,000 data using the bootstrap method. Following the procedure described in the previous paragraph, we obtained the initial squeezing $r$, the optical loss $l$, and the phase diffusion $\Delta$. Figure~\ref{fig:diffusing_parameters} displays the estimated parameters as introducing an additional phase noise $\Delta_e$. The estimated phase diffusion agrees well with the theoretical prediction $\Delta = \sqrt{\Delta_{0}^{2} + \Delta_{e}^{2}}$, which can be derived from the fact that the variance of the sum of independent random variables is simply the sum of the variance of the random variables. The initial squeezing and the loss also behave as expected, exhibiting no changes due to the added noise.


\section{Normally ordered moment method for nonclassicality detection}
Normally ordered moment method~\cite{Agarwal93} employs an $n \times n$ matrix of normally order moments $\mathbf{M}(n)$ as
\begin{equation}
    \mathbf{M}(n) = \begin{bmatrix}
    1 & \langle :\hat{x} : \rangle  & \cdots & \langle :\hat{x}^{n-1} : \rangle \\
    \langle :\hat{x} : \rangle & \langle :\hat{x}^2 : \rangle & \cdots &  \langle :\hat{x}^{n} : \rangle \\
    \vdots & \vdots & \ddots & \vdots \\
    \langle :\hat{x}^{n-1} : \rangle & \langle :\hat{x}^{n} : \rangle & \cdots & \langle :\hat{x}^{2n-2} : \rangle \\
    \end{bmatrix},
    \label{eq:matrix}
\end{equation}
where $\expect{:\hat{x}^{j}:}$ represents the normally ordered moment of the $j$th order. For any classical state, the matrix $\mathbf{M}(n)$ has no negative eigenvalue with all $n \in \{ 2, 3, 4, \cdots \}$. Therefore, if there exists at least one negative eigenvalue of the matrix $\mathbf{M}(n)$ for some $n$, a given state is certified to be nonclassical. We note that the normally ordered moment $\expect{:\hat{x}^{j}:}$ can be rewritten as \cite{Wunsche1999}
\begin{equation}
    \expect{:\hat{x}^{j}:} = \frac{1}{2^{j/2}} \expect{H_{j} ( \hat{x} / \sqrt{2} ) },
\end{equation}
where $H_j(x)$ is the Hermite polynomial of the $j$th degree. It is straightforward to observe that the negative eigenvalue of $\mathbf{M}(2)$ certifies the existence of squeezing. Furthermore, the eigenvalues of $\mathbf{M}(n)$ with $n \geq 3$ generally involve higher-order moments. This is why $\mathbf{M}(n)$ with $n \geq 3$ can detect nonclassical states without squeezing. To use the moment method for nonclassicality detection, the following sampling expression is used:
\begin{equation}
    \langle :\hat{x}^{j} : \rangle \simeq {1 \over 2^{j/2}N} \sum_{i = 1}^{N} H_j(x_i/\sqrt{2}),
\end{equation}
where $x_{i}$ is the $i$-th quadrature measurement, and $N$ is the total number of data.

\section{Entanglement potential of phase-diffused squeezed vacuum}
To determine whether the phase-diffused squeezed vacuum can produce quantum entanglement through a beam-splitter, we calculate the entanglement potential (EP) of the phase-diffused squeezed vacuum in the experiment~\cite{Asboth05}. We first reconstruct the density operator of the phase-diffused squeezed vacuum using the maximum-likelihood estimation method \cite{Lvovsky04}, which is based on quadrature measurement by rotating the quadrature angle. The reconstructed density operator is expressed as
\begin{equation}
    \hat{\sigma} = \sum_{n=0}^{N_c} \sum_{m=0}^{N_c} \sigma_{nm} \ketbra{n}{m},
\end{equation}
where $\vert n\rangle$ is the $n$-photon Fock state, and $N_c$ is the cutoff dimension. Here we choose $N_c = 10$ because the populations in the Fock basis are concentrated in low photon-number Fock states for the case of weekly squeezed vacuum. Then, we calculate the EP of the phase-diffused squeezed vacuum by using the following expression:
\begin{equation} \label{eq:EPD}
    \mathrm{EP}(\hat{\sigma}) = \mathrm{\log}_2 \vert \vert \hat{\rho}_{\hat{\sigma}}^{T_A} \vert \vert,
\end{equation}
where $\hat{\rho}_{\hat{\sigma}} = \hat{U}_{\mathrm{BS}} (\hat{\sigma} \otimes \ketbra{0}{0}) \hat{U}_{\mathrm{BS}}^\dagger$ represents a two-mode quantum state generated by coherently mixing $\hat{\sigma}$ and vacuum $\ketbra{0}{0}$ through a 50:50 beam-splitter $\hat{U}_{\mathrm{BS}}$:
\begin{equation}
    \hat{\rho}_{\hat{\sigma}} = \sum_{n=0}^{N_{c}} \sum_{m=0}^{N_{c}} \sum_{j = 0}^{n} \sum_{k = 0}^{m} \sigma_{nm} \sqrt{\frac{1}{{2^{n+m}}} \binom{n}{j} \binom{m}{k}} \ketbra{j}{k} \otimes \ketbra{n-j}{m-k},
\end{equation}
and the superscript $T_A$ denotes the partial transpose of the density operator, which yields
\begin{equation} \label{eq:PTD}
    \hat{\rho}_{\hat{\sigma}}^{T_{A}} = \sum_{n=0}^{N_{c}} \sum_{m=0}^{N_{c}} \sum_{j = 0}^{n} \sum_{k = 0}^{m} \sigma_{nm} \sqrt{\frac{1}{{2^{n+m}}} \binom{n}{j} \binom{m}{k}} \ketbra{j}{k} \otimes \ketbra{m-k}{n-j}.
\end{equation}
Here $\vert\vert \hat{o} \vert\vert \equiv \mathrm{tr} \sqrt{\hat{o}^{\dag} \hat{o}}$ is the trace norm of the Hermitian operator $\hat{o}$ which equals with the absolute sum of the eigenvalues for $\hat{o}$. Therefore, we can compute EP in Eq.~\eqref{eq:EPD} by solving the eigenvalue problem for the partially transposed density operator in Eq.~\eqref{eq:PTD}.